\documentclass[prd,preprint,showpacs,preprintnumbers,amsmath,amssymb,eqsecnum,nofootinbib,groupedaddress,superscriptaddress,pdflatex]{revtex4-1}
\usepackage{amsfonts,mathrsfs}
\usepackage{bm}
\usepackage{graphicx,color}
\usepackage{hyperref}
\usepackage{ulem}

\begin{document}
\preprint{RUP-24-17}
\preprint{YITP-24-111}
%
\title{Geometrical origin for the compaction function for primordial black hole formation}
\author{Tomohiro Harada}
\email{harada@rikkyo.ac.jp}
\affiliation{Department of Physics, Rikkyo University, Toshima,
Tokyo 171-8501, Japan}
\author{Hayami Iizuka}
\email{h.iizuka@rikkyo.ac.jp}
\affiliation{Department of Physics, Rikkyo University, Toshima,
Tokyo 171-8501, Japan}
\author{Yasutaka Koga}
\email{yasutaka.koga@yukawa.kyoto-u.ac.jp}
\affiliation{Yukawa Institute for Theoretical Physics, Kyoto University, Kyoto 606-8502, Japan}

\author{Chul-Moon Yoo}
\email{yoo.chulmoon.k6@f.mail.nagoya-u.ac.jp}
\affiliation{Graduate School of Science,
Nagoya University, Nagoya 464-8602, Japan}
\begin{abstract}
We propose a geometrical origin for the Shibata-Sasaki compaction function, which is known to be a reliable indicator of primordial black hole formation at least during radiation domination. In the 
long-wavelength limit, we identify it with a compactness function in the static spacetime obtained by removing the cosmological scale factor from the metric and this explains why it cannot be greater than $1/2$. If its maximum is below $1/2$, the perturbation is of type I. If its maximum equals $1/2$, it corresponds to an extremal surface, which is simultaneously a bifurcating trapping horizon and admits a circular photon orbit in the static spacetime. In the long-wavelength regime of the physical expanding Universe, the Shibata-Sasaki compaction reaches its maximum value of $1/2$ at maximal and minimal surfaces on the constant time spacelike hypersurface, which feature a type II perturbation and both correspond to photon spheres expanding along with the cosmological expansion. Thus, the Shibata-Sasaki compaction measures how close to the type II configuration the perturbed region is.
\end{abstract}

\date{\today}

\maketitle

\tableofcontents

\newpage

\section{Introduction}

Recently, primordial black holes (PBHs) have been the subject of intensive investigation, driven by observational and theoretical advancements across various related fields of research. This surge in interest was sparked by the discovery of binary black holes with masses of around 30 solar masses via gravitational waves in 2015~\cite{LIGOScientific:2016aoc}. Further studies have indicated that there remains a window at asteroid-scale masses, within which PBHs could account for all the dark matter in the Universe (see~\cite{Carr:2020gox} and references therein). To theoretically predict the abundance and other physical properties of PBHs, a detailed understanding of the physics governing their formation processes is crucial.

One of the standard formation scenarios is the direct collapse of primordial perturbations generated during inflation. Quantum fluctuations are produced and stretched in the inflationary phase. After inflation and reheating, spacetime with a nonlinearly large amplitude of superhorizon perturbations can be described by long-wavelength solutions in the radiation-dominated era. These long-wavelength solutions with adiabatic perturbations are determined by a single function of spatial coordinates that specifies the curvature perturbation. If the amplitude of the perturbation is sufficiently large, it collapses into a black hole after the horizon entry. Since the formation process involves highly nonlinear general relativistic hydrodynamics, numerical relativity simulations are typically required, even in spherical symmetry.

Because of the high cost of such computations, simple physical and/or phenomenological thresholds for PBH formation have been sought. In particular, the so-called compaction function (or just compaction) has played a central role in this context. The compaction function  was originally introduced by Shibata and Sasaki (1999)~\cite{Shibata:1999zs} on the constant-mean-curvature slice and was empirically shown to give a threshold of PBH formation during radiation domination when its value is approximately 0.4. This function has been reformulated in terms of curvature perturbation in Refs.~\cite{Harada:2015yda,Harada:2023ffo}.

We refer to this function as the Shibata-Sasaki compaction 
${\mathscr C}_{\rm SS}$ to distinguish it from the legitimate compaction ${\mathscr C}_{\rm com}=(\delta M/R)_{\rm com}$ later defined in the comoving slice in~\cite{Harada:2015yda,Musco:2018rwt,Escriva:2019phb,Harada:2023ffo}, where $\delta M$ and $R$ are the Misner-Sharp mass excess and the areal radius, respectively. These two compactions are related to each other through a constant factor for the equation of state $p=w\rho$ but not for more general cases.
Both of the compactions give a measure of the early perturbations that may later lead to the formation of black holes.

 Despite the importance of the Shibata-Sasaki compaction for PBH formation, we have not yet identified any direct interpretation of this function. On the other hand, threshold studies in many numerical simulations have been discussed in terms of ${\mathscr C}_{\rm com}$~\cite{Musco:2018rwt,Escriva:2019phb,Musco:2020jjb}. In this paper, we propose a geometrical origin for the Shibata-Sasaki compaction as a {\it compactness} function ${\cal C}=M/R$, where $M$ is the Misner-Sharp mass, in the static spacetime obtained through a conformal transformation from the cosmological long-wavelength solutions. The compactness ${\cal C}$ is directly related to null expansions and, consequently, trapping horizons, whereas neither the compaction ${\mathscr C}_{\rm SS}$ nor ${\mathscr C}_{\rm com}$ exhibits such a relationship. It is interesting to note that this interpretation illuminates a type II perturbation, which has recently been shown to produce a new type of PBHs in radiation domination~\cite{Kopp:2010sh,Uehara:2024yyp}, admitting circular photon orbits in the static spacetime. The present work is inspired by the recent report on a certain numerical coincidence in the thresholds between PBH formation and circular photon orbits in Ref.~\cite{Ianniccari:2024eza}, whereas the analysis conducted here is entirely independent of it.

This paper is organised as follows. In Sec.~\ref{sec:compactness}, we 
introduce the compactness function and 
the conformal transformation in general spherically symmetric spacetimes. In Sec.~\ref{sec:interpretation}, we introduce a conformal compactness function 
and identify it with the Shibata-Sasaki compaction in the long-wavelength limit. In Sec.~\ref{sec:circular}, we demonstrate that the Shibata-Sasaki compaction identifies circular photon orbits in the unphysical static spacetime and, consequently, type II perturbations.  
Section~\ref{sec:conclusion} concludes the paper.
We use units in which $G=c=1$ and the sign convention in 
Wald~\cite{Wald:1984rg}.

\section{Compactness function and conformal transformation}
\label{sec:compactness}
Let $({\cal M}, g)$ be a spherically symmetric spacetime in four dimensions.
The line element in this spacetime can be 
written in the following form: 
\begin{equation}
 ds^{2}=g_{AB}(x^{C})dx^{A}dx^{B}+R^{2}(x^{C})d\Omega^{2},
\end{equation}
where $x^{A}$ ($A \cdots=0,1$) are coordinates in the two-dimensional 
spacetime ${\mathcal M}^{2}$ with a metric $g_{AB}dx^{A}dx^{B}$ 
and $x^{a}$ ($a\cdots=2,3$) are coordinates in the unit two-sphere with the metric 
$d\Omega^{2}=\sigma_{ab}dx^{a}dx^{b}$, while we use $x^{\mu}$ ($\mu=0,1,2,3$) for coordinates in the four-dimensional spacetime.
The Misner-Sharp mass, or the Kodama mass, is written as
~\cite{Misner:1964je,Kodama:1979vn}
\begin{equation}
 M(x^{C})=\frac{1}{2}R(x^{C})\left[
1-g^{AB}\nabla_{A}R(x^{C})\nabla_{B}R(x^{C})\right].
\end{equation}
Note that this quantity is gauge independent in the sense that $M(x^{A})$ is a scalar 
against a coordinate transformation on ${\mathcal M}^{2}$ spanned by $x^{A}$.
In literature, $M/R$ is often called compactness, so  
we define a compactness function (or just compactness) $C(x^{A})$ as 
the ratio of the Misner-Sharp mass $M(x^{A})$ to the areal radius $R(x^{A})$, i.e., 
\begin{equation}
 C(x^{A})=\frac{M(x^{A})}{R(x^{A})}.
\end{equation}
This is also gauge independent in the above sense. Note that the compactness is closely related to the null expansions $\theta_{+}$ and $\theta_{-}$
as follows:
\begin{equation}
 C(x^{A})=\frac{1}{2}\left[1+\frac{1}{2}R^{2}\theta_{+}\theta_{-}\right],
\end{equation}
where 
\begin{equation}
 \theta_{\pm }:=l^{A}_{\pm }\nabla_{A}\ln R^{2}
\end{equation}
are the radial null expansions
along with the future-pointing radial null vectors 
$l_{\pm}^{A}$ with $l_{+}^{A}l_{-A}=-1$. 
Therefore, we have $C=1/2$ on the trapping horizon, which is a hypersurface 
foliated by marginally trapped spheres with $\theta_{+}\theta_{-}=0$, 
while $C>(<)1/2$ on the (un)trapped spheres with 
$\theta_{+}\theta_{-}>(<)0$~\cite{Hayward:1993wb}.

Then, let us rewrite the metric as 
\begin{eqnarray}
 ds^{2}=\Omega^{2}(x^{C})\left[
\tilde{g}_{AB}(x^{C})dx^{A}dx^{B}+\tilde{R}^{2}(x^{C})d\Omega^{2}\right] 
= \Omega^{2}(x^{C})d\tilde{s}^{2},
\end{eqnarray}
where 
\begin{eqnarray}
 g_{AB}=\Omega^{2}\tilde{g}_{AB}, \quad R= \Omega \tilde{R}.
\end{eqnarray}
The spacetime defined by $({\cal M},\tilde{g})$ is called an ``unphysical'' spacetime~\cite{Wald:1984rg}, although it does not mean that this has nothing to do with the ``physical'' spacetime.
Then, we can define the compactness in the unphysical spacetime as
\begin{equation}
 \tilde{C}(x^{C})=\frac{\tilde{M}(x^{C})}{\tilde{R}(x^{C})}
=\frac{1}{2}\left[1-\tilde{g}^{AB}\tilde{\nabla}_{A}\tilde{R}(x^{C})
\tilde{\nabla}_{B}\tilde{R}(x^{C})\right].
\label{eq:conformal_compactness}
\end{equation}
By this definition, we can derive the following relation between $C$ and $\tilde{C}$
\begin{equation}
 \tilde{C}=C+\frac{1}{2}g^{AB}\nabla_{A}(R^{2})\nabla_{B}\ln \Omega-
\frac{1}{2}R^{2}g^{AB}\nabla_{A}\ln \Omega \nabla_{B}\ln \Omega
\label{eq:CtildefromC}
\end{equation}
or
\begin{equation}
 C=\tilde{C}-\frac{1}{2}\tilde{g}^{AB}\tilde{\nabla}_{A}(\tilde{R}^{2})\tilde{\nabla}_{B}\ln \Omega-\frac{1}{2}\tilde{R}^{2}\tilde{g}^{AB}\tilde{\nabla}_{A}\ln \Omega \tilde{\nabla}_{B}\ln \Omega.
\label{eq:CfromCtilde}
\end{equation}

The compactness in the unphysical spacetime is straightforward to obtain. Given a metric and a conformal factor $\Omega$, we can immediately calculate it.
It does not refer to the matter contents, whether they are a single perfect fluid, multiple perfect fluids, scalar fields or any other fields and/or the gravitational theory that governs the relation between the metric and the matter fields.

\section{Geometrical origin for the Shibata-Sasaki compaction function}
\label{sec:interpretation}
The metric in the cosmological conformal decomposition is written in the following form:
\begin{equation}
 ds^{2}=a^{2}(\eta)\left[-{\alpha}^{2}d\eta^{2}+\psi^{4}
\tilde{\gamma}_{ij}(dx^{i}+\beta^{i}d\eta)(dx^{j}+\beta^{j}d\eta)\right],
\end{equation}
where $\psi$, $\tilde{\gamma}_{ij}$, $\alpha$ and $\beta^{i}$ are functions of $x^{\mu}=(t,{\bm x})$ and we fix the normalisation of $\tilde{\gamma}_{ij}$ so that $\det (\tilde{\gamma}_{ij})=\det (\eta_{ij})$ with $\eta_{ij}=\eta_{ij}({\bm x})$ being the static metric for the flat three-space. For later convenience, we have introduced the conformal time $\eta$ and the scale factor $a(\eta)$, where $\eta$ is related to the cosmological time $t$ through $a(\eta)d\eta=dt$ and $a(\eta)$ is the scale factor of the background Friedmann-Lema\^{i}tre-Robertson-Walker (FLRW) solution with the flat spatial curvature.
Rewriting $\alpha$ and $\psi$ as  
\begin{equation}
 \alpha=1+\chi, \psi=\Psi(1+\xi), \tilde{\gamma}_{ij}=\eta_{ij}+h_{ij},
\end{equation}
and taking appropriate gauge conditions, 
we can construct the so-called cosmological long-wavelength 
solutions of the Einstein equation, admitting the following power series~\cite{Shibata:1999zs,Lyth:2004gb,Harada:2015yda}:
\begin{equation}
 \Psi=\Psi({\bm x})=O(1), \chi(\eta,{\bm x})=O(\epsilon^{2}), \beta^{i}(\eta,{\bm x})=O(\epsilon), \xi(\eta,{\bm x})=O(\epsilon^{2}), h_{ij}(\eta,{\bm x})=O(\epsilon^{2}),
 \label{eq:power_series}
\end{equation}
where $\epsilon=k/(aH_{b})$ is a small parameter for the gradient expansion with $H_{b}:=(da/d\eta)/a^{2}$ being the Hubble parameter of the background flat FLRW solution
\footnote{In the presence of nonperturbative isocurvature perturbation, we cannot generally assume Eq.~(\ref{eq:power_series}) as indicated in Ref.~\cite{Lyth:2004gb}. 
Here we simply assume
that isocurvature perturbation appears only from $O(\epsilon^{2})$ in the metric,
if any, in the long-wavelength regime.}.
In particular, the zeroth-order quantity $\Psi({\bm x})$
generates the long-wavelength solutions
and is independent of the choice of time slicing
as long as it admits the above scheme of long-wavelength solutions,
including the comoving slice, the uniform density slice and 
the constant-mean-curvature (CMC) slice.
$\zeta:=(1/2)\ln \Psi$ is called curvature perturbation in cosmology.

In spherically symmetric spacetimes, we can write the line element in the following form
\begin{equation}
 ds^{2}=a^{2}(\eta)\left\{-{\alpha^{2}(\eta,r)}d\eta^{2}+\psi^{4}(\eta,r)\left[
e^{2\lambda(\eta,r)}(dr+\beta^{r}(\eta,r)d\eta)^{2}+e^{-\lambda(\eta,r)}r^{2}d\Omega^{2}\right]\right\},
\end{equation}
where $\lambda=O(\epsilon^2)$ is introduced in place of $h_{ij}$. 
Then, using Eq.~(\ref{eq:conformal_compactness}) with $\Omega=a(\eta)$, 
we obtain the compactness in the unphysical spacetime, which 
we will hereafter call the conformal compactness, as
\begin{eqnarray}
 \tilde{C}&=&\frac{1}{2}
\left\{
1+\alpha^{-2}r^{2}[\partial_{\eta}(\psi^{2}e^{-\lambda/2})]^{2}-2\alpha^{-2}\beta^{r}r\partial_{\eta}(\psi^{2}e^{-\lambda/2})\partial_{r}(\psi^{2}e^{-\lambda/2}r) \right. \nonumber \\
&& \left. -[\psi^{-4}e^{-2\lambda}-\alpha^{-2}(\beta^{r})^{2}][\partial_{r}(\psi^{2}e^{-\lambda/2}r)]^{2}
\right\}.
\end{eqnarray}
Taking the limit $\epsilon\to 0$ in the above, we obtain
\begin{equation}
 \tilde{C}(r)\approx \frac{1}{2}\left\{1-\Psi^{-4}[\partial_{r}(\Psi^{2}r)]^{2}\right\},
\label{eq:conformal_compactness_limit}
\end{equation}
where the weak equality denotes the equality in the limit $\epsilon\to 0$.
This result does not depend on the slicing condition as long as it admits 
the long-wavelength scheme.

In fact, the line element in the zeroth-order of long-wavelength solutions in spherical symmetry is written in the following form
\begin{equation}
ds^{2}\approx a^{2}(\eta)\left[-d\eta^{2}+\Psi^{4}(r)(dr^{2}+r^{2}d\Omega^{2})\right].
\label{eq:zeroth_spherical_solution}
\end{equation}
Thus, the physical spacetime is conformal to the (ultra)static unphysical spacetime with the metric $\tilde{g}_{\mu\nu}$ in the 
zeroth-order long-wavelength solutions.
We can recover Eq.~(\ref{eq:conformal_compactness_limit})
by calculating the compactness directly using Eq.~(\ref{eq:conformal_compactness}) 
for the line element (\ref{eq:zeroth_spherical_solution}). 
\footnote{On the other hand, the long-wavelength limit of the relation (\ref{eq:CfromCtilde})
is not so obvious because of the nontrivial contribution due to the nonvanishing shift vector $\beta^{i}$.}

The Shibata-Sasaki compaction function ${\mathscr C}_{\rm SS}$ is defined 
as~\cite{Shibata:1999zs}
\begin{equation}
{\mathscr C}_{\rm SS}=\left(\frac{1}{R}\int dr
4\pi R^{2}(\partial_{r}R)  \delta \rho\right)_{\rm CMC},
\end{equation}
where the integral on the right-hand side is that 
on the constant time spacelike hypersurface in the background flat FLRW spacetime and ``CMC'' implies that it is evaluated in the CMC slice. Note that in this paper the compaction and the compactness are distinguished from each other by the calligraphic ${\mathscr C}$ and the italic $C$, respectively.
As is shown in Ref.~\cite{Harada:2023ffo}, 
despite the initial naive intention, it has turned out that this is not equal to the ``legitimate'' compaction ${\mathscr C}_{\rm CMC}$, that is, 
\begin{equation}
{\mathscr C}_{\rm SS}\neq {\mathscr C}_{\rm CMC}:=\left(\frac{\delta M}{R}\right)_{\rm CMC},
\end{equation}
where $\delta M$ is the excess in the Misner-Sharp mass $M$. This is  
because the velocity perturbation gives nonvanishing contribution to $\delta M$ in the CMC slice.
On the other hand, we can show that using the density perturbation in the CMC slice in the long-wavelength solution
\begin{equation}
\delta_{\rm CMC}:=\frac{\delta \rho_{\rm CMC}}{\rho_{b}}\approx 
-\frac{4}{3}\left(\frac{1}{aH_{b}}\right)^{2}\frac{1}{\Psi^{5}}
\frac{1}{r^{2}}\partial_{r}\left(r^{2}\partial_{r}\Psi\right),
\end{equation}
where $\rho_{b}$ is the energy density in the background flat FLRW solution,  
the Shibata-Sasaki compaction ${\mathscr C}_{\rm SS}(r)$ 
admits the following expression in the 
long-wavelength limit~\cite{Harada:2015yda} \footnote{On the comoving slice, we can easily show 
\begin{equation*}
{\mathscr C}_{\rm com}:=\left(\frac{\delta M}{R}\right)_{\rm com}=
\left(\frac{1}{R}\int dr 4\pi R^{2}(\partial_{r}R)  \delta \rho\right)_{\rm com}.
\end{equation*}
In the long-wavelength limit, if the matter field is given by a perfect fluid with the equation of state $p=w\rho$, where $p$, $\rho$ and $w$ are the pressure, the energy density and a constant, respectively, the relation between 
${\mathscr C}_{\rm SS}$ and ${\mathscr C}_{\rm com}$ is given by~\cite{Harada:2015yda} 
\begin{equation*}
{\mathscr C}_{\rm SS}\approx \frac{5+3w}{3(1+w)}{\mathscr C}_{\rm com},
\end{equation*}
but in more general cases, this is not the case.
}
\begin{equation}
  {\mathscr C}_{\rm SS}(r)\approx \frac{1}{2}\left\{1-\Psi^{-4}\left[\partial_{r}(\Psi^{2}r)\right]^{2}\right\}.
\label{eq:CSS_zeroth}
\end{equation}

From Eqs.~(\ref{eq:conformal_compactness_limit}) and (\ref{eq:CSS_zeroth}), 
we can conclude 
\begin{equation}
 {\mathscr C}_{\rm SS}(r)\approx \tilde{C}(r).
\end{equation}
Therefore, in the long-wavelength limit, the Shibata-Sasaki compaction equals to the compactness  
of the unphysical static spacetime, or the conformal compactness.

The zeroth-order long-wavelength solution~(\ref{eq:zeroth_spherical_solution})
is written in the spatially conformally flat coordinates $(r,x^{2},x^{3})$. If we transform them 
to the areal radial coordinates $(\bar{r},x^{2},x^{3})$ for the unphysical static metric with $\bar{r}=\Psi^{2}(r)r$, we obtain the following 
more familiar form:
\begin{eqnarray}
 ds^{2}&\approx& a^{2}(\eta)\left[-d\eta^{2}+
\frac{d\bar{r}^{2}}{1-2{\mathscr C}_{\rm SS}(r)}+\bar{r}^{2}d\Omega^{2}\right].
\end{eqnarray}
This can be also understood as the asymptotic quasihomogeneous solutions 
developed in the Misner-Sharp formulation through 
\begin{equation}
2{\mathscr C}_{\rm SS}(r)\approx K(\bar{r})\bar{r}^{2},
\end{equation}
where $K(\bar{r})$ is called the curvature profile in the quasihomogeneous solution formulation~\cite{Polnarev:2006aa,Harada:2015yda}.
However, these coordinates cannot cover an extremal surface, which corresponds to coordinate singularity.
To circumvent this drawback, hereafter, we will use the spatially flat coordinates rather than the areal radial coordinates.

Although the Shibata-Sasaki compaction is empirically known to 
be a robust indicator of PBH formation for a perfect fluid with the equation of state $p=w\rho$, we do not know how we can extend this function to more general systems such as those with more general matter fields and/or in modified theories of gravity.
We propose that the conformal compactness not only gives 
a geometrical origin for the Shibata-Sasaki compaction 
but also immediately extends to the systems with more general matter fields and/or in modified gravity, although whether or not it is 
useful as an indicator for the PBH formation is yet unclear.
The conformal compactness is straightforward to be defined  
in the context of the formation of PBHs and other structures.
It is conceptually simple and its geometrical meaning is clear as 
it can describe the trapping of 
photons in the unphysical static spacetime obtained by removing the cosmological expansion from the physical spacetime.

\section{Trapping horizons, circular photon orbits and extremal surfaces}
\label{sec:circular}

We can probe the spacetime geometry with null geodesics.
The trajectories of null geodesic equations are invariant under the conformal transformation, for which the affine parameter $\lambda$ in the physical spacetime is replaced by that in the unphysical 
spacetime $\tilde{\lambda}$ with
\begin{equation}
\frac{d\tilde{\lambda}}{d\lambda}=\frac{c}{\Omega^{2}},
\label{eq:lambdatilde_lambda} 
\end{equation}
where $c$ is a nonzero constant.
In other words, null geodesics follow the same trajectories on 
the physical spacetime and the unphysical spacetime with the different affine parameters $\lambda$ and $\tilde{\lambda}$, 
respectively.
On the other hand, since the signs of the radial null expansions are 
affected by the conformal transformation, 
trapping horizons and trapped regions are also affected. In fact, we have
\begin{equation}
 \tilde{\theta}_{\pm}=\Omega^{-1}[\theta_{\pm}-l^{A}_{\pm}\nabla_{A}\ln \Omega^{2}],
\label{eq:thetatilde_theta}
\end{equation}
where we have defined $\tilde{\theta}_{\pm}$ 
in terms of the radial null vectors
$ \tilde{l}_{\pm}^{A}=\Omega^{-1}l_{\pm}^{A}$
as
\begin{eqnarray}
 \tilde{\theta}_{\pm}=\tilde{l}^{A}_{\pm}\tilde{\nabla}_{A}\ln \tilde{R}^{2}.
\label{eq:thetatilde}
\end{eqnarray}
Equation~(\ref{eq:thetatilde_theta}) implies that the increasing conformal factor along the null vector $l^{A}_{\pm}$ gives a negative term in the null expansion in the unphysical spacetime in comparison to that in the physical spacetime. 
Using $\tilde{\theta}_{\pm}$, of course, we have 
\begin{equation}
 \tilde{C}(x^{A})=\frac{1}{2}\left[1+\frac{1}{2}\tilde{R}^{2}\tilde{\theta}_{+}\tilde{\theta}_{-}\right].
\label{eq:Ctilde}
\end{equation}
So, the conformal compactness is useful to know the
trapping horizons and trapped regions in the unphysical spacetime.

Now, let us consider null geodesics in the unphysical static spacetime 
\begin{equation}
\widetilde{ds}^{2}= -d\eta^{2}+\Psi^{4}(r)(dr^{2}+r^{2}d\Omega^{2}),
\label{eq:unphysical_static_spacetime}
\end{equation}
which is extracted from 
the zeroth-order long-wavelength solution. 
Here, we derive the important properties of the Shibata-Sasaki compaction.
We can take the coordinate components of the radial null vectors as 
\begin{equation}
\tilde{l}^{A}_{\pm}=\frac{1}{\sqrt{2}}(1,\pm \Psi^{-2}).
\end{equation}
This implies 
\begin{equation}
\tilde{\theta}_{\pm}=\pm \sqrt{2}\Psi^{-2}[\ln (\Psi^{2}r)]'=\pm \sqrt{2}\Psi^{-4}r^{-1}F(r),
\label{eq:thetatilde_static}
\end{equation}
where the prime denotes the derivative with respect to $r$ and 
\begin{equation}
    F(r):=(\Psi^{2} r)'=\tilde{R}'(r).
\label{eq:def_F}
\end{equation}
Therefore, there is no trapped sphere in the static spacetime.
From Eqs.~(\ref{eq:Ctilde}), (\ref{eq:thetatilde_static}) and (\ref{eq:def_F}), we can deduce ${\mathscr C}_{\rm SS}(r)=\tilde{C}(r)\le 1/2$, where the equality holds only for $\tilde{\theta}_{+}=\tilde{\theta}_{-}=0$ corresponding to a bifurcating trapping horizon~\cite{Maeda:2009tk}. Since $R=a(\eta)\tilde{R}(r)=a(\eta)\Psi^{2}(r)r$, 
this also corresponds to an extremal surface on the constant $\eta$ spacelike hypersurface in the physical spacetime. These results are the direct consequence of the fact that the Shibata-Sasaki compaction is the compactness of the unphysical static spacetime.

The null geodesic has the conserved quantities 
\begin{eqnarray}
 E=\dot{\eta}, \quad 
 L=\Psi^{4}r^{2}\dot{\phi}, 
\end{eqnarray}
where the dot denotes the ordinary differentiation with respect to $\tilde{\lambda}$.
The radial motion can be written in the form
\begin{equation}
 \dot{r}^{2}+V(r)=0; \quad
 V(r)=\Psi^{-4}\left(-1+\frac{b^{2}}{\Psi^{4}r^{2}}\right), 
\end{equation}
where $E$ is absorbed by rescaling the affine parameter and $b:=L/E$ is the impact parameter. If we assume $b\ne 0$, $\Psi\to \Psi_{0}(>0)$ as $r\to 0$ and $\Psi\to 1$ as $r\to \infty$, we have $V\approx   b^{2}/(\Psi_{0}^{8}r^{2})$ as $r\to 0$ and $V\to -1$ as $r\to \infty$.

\begin{figure}
\begin{center}
\includegraphics[width=0.45\textwidth]{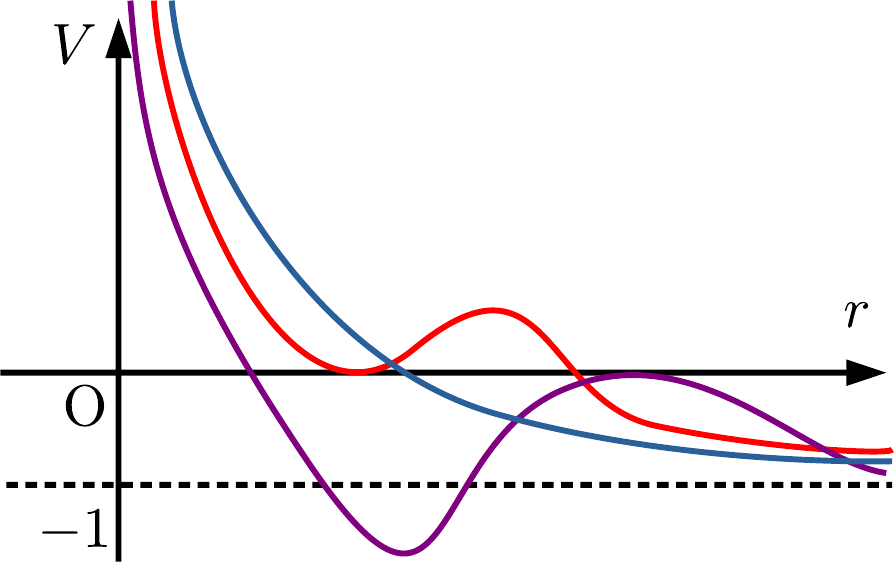}
\caption{Schematic figure for the shapes of the effective potentials. The blue, red and purple curves show the potentials with no circular orbit, a stable circular orbit and an unstable circular orbit, respectively. The radius of the circular photon orbit is given by a double root of $V(r)$. \label{fig:effective_potential}}
\end{center}
\end{figure}

There exists a circular null geodesic in the static unphysical spacetime
if and only if there is $r_{p}(>0)$ such that $V(r_{p})=V'(r_{p})=0$.
Since the trajectories of null geodesics are not affected by the conformal transformation, this circular photon orbit in the unphysical spacetime corresponds to that staying at $r=r_{p}$ 
but with the expanding areal radius $a(\eta)r_{p}$ in the physical spacetime.
Since 
\begin{eqnarray}
 V'(r)=(\Psi^{-4})'\Psi^{4}V(r)-2\Psi^{-4}b^{2}(\Psi^{2}r)^{-3}(\Psi^{2}r)',
 \label{eq:V'}
\end{eqnarray}
the condition $V=V'=0$ is equivalent to the following
\begin{eqnarray}
 (\Psi^{2}r)'=0, \quad
 b^{2}= \Psi^{4}r^{2}
 \label{eq:circular_orbit}
\end{eqnarray}
at $r=r_{p}$. The first equation gives the radius of the circular photon 
orbit $r_{p}$, while the second gives the impact parameter $b=b_{p}$ of the photon. 
Thus, we can reformulate the problem as finding the zero 
$r_{p}$ of $F(r)=(\Psi^{2}r)'$
and then calculating the impact parameter $b_{p}$ using 
Eq.~(\ref{eq:circular_orbit}). 
Equations~(\ref{eq:V'}) and (\ref{eq:circular_orbit}) imply
\begin{eqnarray}
V''(r_{p})=- \frac{2F'(r_{p})}{\Psi^{6}(r_{p})r_{p}}.
\label{eq:V''}
\end{eqnarray}
Figure~\ref{fig:effective_potential} schematically exhibits the shape of the effective potentials corresponding to those with no circular orbit, a stable circular orbit and an unstable circular orbit with blue, red and purple curves, respectively.

Then, from Eq.~(\ref{eq:CSS_zeroth}), 
we can conclude 
\begin{equation}
 \tilde{C}(r_{p})={\mathscr C}_{\rm SS}(r_{p})=\frac{1}{2}.
\end{equation}
As a corollary, the circular orbit at $r=r_{p}$ 
corresponds to a trapping horizon in the unphysical static spacetime. 
This also implies that if there is a circular photon orbit $r=r_{p}$, 
we have $\tilde{R}'=0$ there corresponding to an extremal surface,
and vice versa.
In the physical spacetime, this implies that if and only if there is an extremal surface on the constant time hypersurface, there is a null geodesic staying on the extremal surface during the cosmological expansion.

In the trivial case $\Psi(r)=1$, we have $F(r)=1$.
Assuming $\Psi(r)\to \Psi_{0}>0$ as $r=0$, we have $F(r)\to \Psi_{0}^{2}$ as $r\to 0$. If we further assume $\Psi(r)$ approaches 1
as $r\to \infty$, we have $F(r)\to 1$ as $r\to \infty$. 
If $F(r)$ is everywhere positive, $\tilde{C}={\mathscr C}_{\rm SS}<1/2$ everywhere and hence the whole static 
spacetime is untrapped.

Let us continuously deform $F(r)$ from the trivial one 
keeping the boundary conditions at $r\to 0$ and $r\to \infty$ and 
focus on the zeros of $F(r)$. From Eq.~(\ref{eq:def_F}), we can reconstruct $\Psi(r)$ from $F(r)$ by direct integration
\begin{equation}
\Psi^{2}(r)=\frac{1}{r}\int^{r}_{0}F(\tilde{r})d\tilde{r}.
\end{equation}
Thus, it turns out that however large $\Psi_{0}$ is, we can always reconstruct $\Psi(r)$ such that $F(r)$ is positive everywhere.
This implies that the large value of $\Psi_{0}$ alone is not enough to 
guarantee the existence of extremal surfaces.
This is consistent with the numerical result in Ref.~\cite{Shimada:2024eec}.

As we continuously deform $F(r)$,
we will generically first encounter a double zero $r_{c}$
of $F(r)$ with $F'(r_{c})=0$ when $F(r)$ turns to get zeros.
This double zero $r=r_{c}$ corresponds to a marginally stable circular orbit because Eq.~(\ref{eq:V''}) 
implies $V''(r_{c})=0$.
In this critical case,
the area of the sphere of constant $r$ takes an inflection point at $r=r_{c}$ with $\tilde{R}'(r_{c})=\tilde{R}''(r_{c})=0$, while the 
area is monotonically increasing with $r$.
This is also the case for the constant $\eta$ spacelike hypersurface in the physical spacetime because the conformal factor $\Omega=a(\eta)$ is only a constant.
In this case, 
the unphysical spacetime is untrapped for $0<r<r_{c}$ and $r_{c}<r$, while there is a 
bifurcating trapping horizon at $r=r_{c}$. 

If we continuously deform $F(r)$ further, $F(r)$ will generically get to have 
two zeros $r_{p \pm}$ with $r_{p+}> r_{p-}$, with $F'(r_{p+})>0$ and $F'(r_{p-})<0$. Since $V(r_{p\pm })=V'(r_{p\pm })=F(r_{p\pm})=0$ by construction, 
Eq.~(\ref{eq:V''}) implies $V''(r_{p+})<0$ and $V''(r_{p-})>0$. That is, $r=r_{p+}$ and $r=r_{p-}$ correspond to an unstable circular photon orbit and a 
stable circular photon orbit, 
respectively.
In this case, $\tilde{C}(r)={\mathscr C}_{\rm SS}(r)$ takes two maximum values of $1/2$ at $r=r_{p\pm }$, while
it is smaller than $1/2$ for $0<r<r_{p-}$, $r_{p-}<r<r_{p+}$ and $r_{p+}<r$ as seen in Eq.~(\ref{eq:CSS_zeroth}).
The sphere of constant $(\eta,r)$ is untrapped for all of $0<r<r_{p-}$, 
$r_{p-}<r<r_{p+}$ and $r_{p+}<r$ and marginally trapped at $r=r_{p-}$ and $r=r_{p+}$. 
There are two bifurcating trapping horizons in this case, the one at $r=r_{p-}$
and the other at $r=r_{p+}$.
In fact, since $\tilde{R}=\Psi^{2}r$, we can find that 
$r=r_{p-}$ and $r=r_{p+}$ correspond to a maximal surface and a
minimal surface, respectively, on the spacelike 
hypersurface of constant $\eta$.
The area of the sphere monotonically increases for $0<r<r_{p-}$, 
takes a maximum at $r=r_{p-}$, monotonically decreases for $r_{p-}<r<r_{p+}$, takes a minimum at $r=r_{p+}$ and monotonically increases again for $r_{p+}<r$. The spatial geometry on the spacelike hypersurface of constant $\eta$ 
in the physical spacetime has the same property. 
The sets of initial data with this property are called type II perturbations, while those with a monotonically increasing areal radius are called type I perturbations~\cite{Kopp:2010sh}.
So, the type II perturbation admits photon orbits staying at $r=r_{p+}$ and $r=r_{p-}$ during the cosmological expansion.
This implies that the maximal and minimal surfaces of type II configurations both correspond to expanding photon spheres, or more precisely ``photon surfaces''~\cite{Claudel:2000yi,Koga:2019uqd} in the physical expanding spacetime. 
The photon spheres at the maximal and minimal surfaces in the unphysical static spacetime are stable and unstable 
as we have seen from the stability of circular null geodesics in the above.
See Appendix~\ref{sec:stability} for the definition of photon surfaces and their stability and the stability of the expanding photon spheres obtained here. It turns out that the stable photon sphere at the maximal surface remains stable in the physical expanding spacetime under the null energy condition, while the unstable photon sphere at the minimal surface may be stabilised by the cosmological expansion.
The above discussion is consistent with~\cite{Uehara:2024yyp}.
We can easily extend the discussion to 
further deformation of $F(r)$, 
in which $F(r)$ may take more than two zeros.
On the other hand, as we have shown, however large $\Psi_{0}$ is, we
can always construct $\Psi(r)$ such that $F(r)$ is everywhere positive. This proves that there exists a type I perturbation, however large $\Psi_{0}$ is.

\begin{figure}
\begin{center}
\begin{tabular}{cc}
\includegraphics[width=0.48\textwidth]{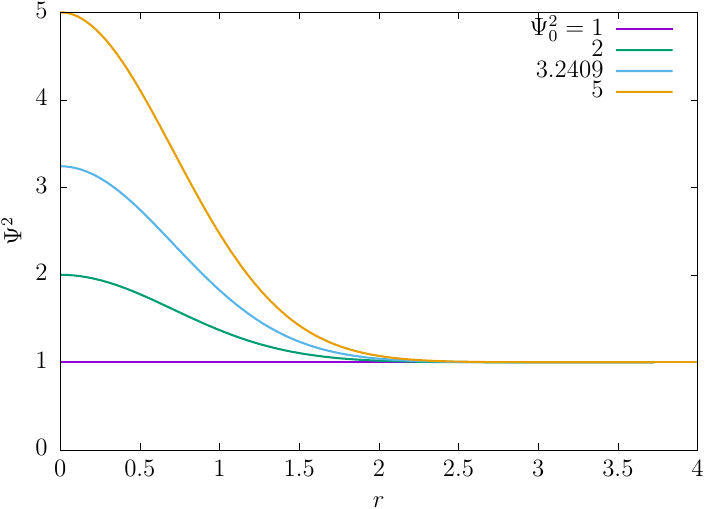} &
\includegraphics[width=0.48\textwidth]{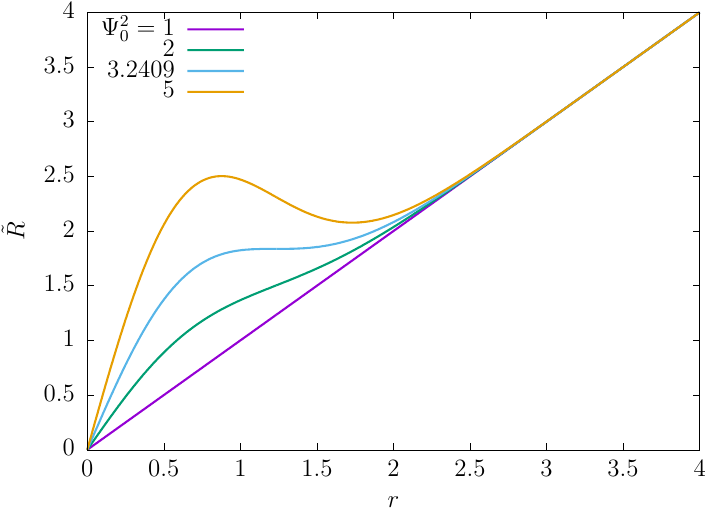} \\
\includegraphics[width=0.48\textwidth]{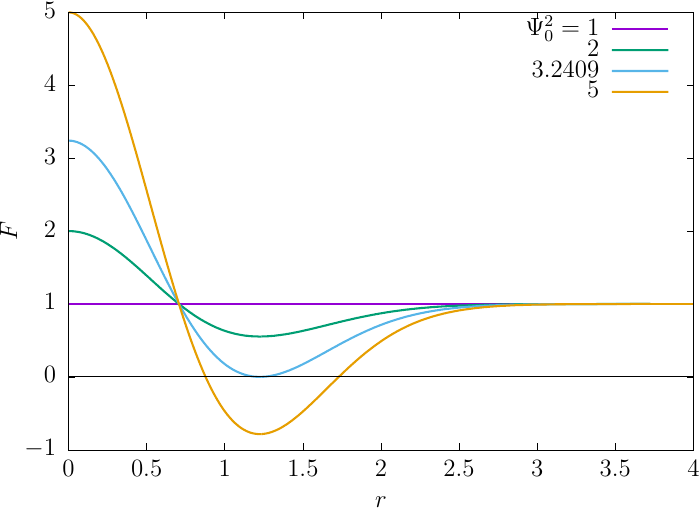} & 
\includegraphics[width=0.48\textwidth]{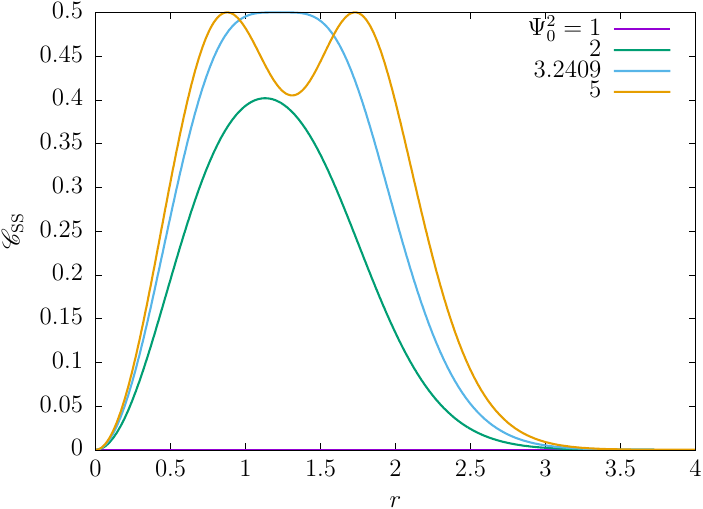}
\end{tabular}
\caption{The shape of functions $\Psi(r)$, $\tilde{R}(r)$, $F(r)$ and ${\mathscr C}_{\rm SS}(r)$ for $\Psi(r)$ given by Eq.~(\ref{eq:Gaussian_Psi}). The purple, green, cyan and yellow curves stand for the cases with $\Psi_{0}^{2}=1$, 
$2$, $3.2409$ and $5$, respectively, all with $\sigma=1$. For $\Psi_{0}^{2}=1$, the geometry is trivial Euclidean, while as we choose $\Psi_{0}^{2}=2$, $3.2409$ and $5$, $F(r)$ is deformed to have no zero, a double zero and two distinct zeros, corresponding to type I, marginal and type II perturbations, respectively.
\label{fig:functions}}
\end{center}
\end{figure}

Although the above proof is complete, it would be helpful for understanding the result to provide examples. Figure~\ref{fig:functions}
shows the shape of functions $\Psi(r)$, $\tilde{R}(r)$, $F(r)$ and ${\mathscr C}_{\rm SS}(r)$ on the top left, top right, bottom left and bottom right panels, respectively, with the Gaussian profile
\begin{equation}
\Psi^{2}=(\Psi_{0}^{2}-1)e^{-r^{2}/\sigma^{2}}+1.
\label{eq:Gaussian_Psi}
\end{equation}
The purple, green, cyan and yellow curves stand for the cases with $\Psi_{0}^{2}=1$, 
$2$, $3.2409$ and $5$, respectively, all with $\sigma=1$. For $\Psi_{0}^{2}=1$, the geometry is trivial Euclidean, while as we take $\Psi_{0}^{2}=2$, $3.2409$ and $5$, $F(r)$ is deformed to have no zero, a double zero and two distinct zeros, i.e., type I, marginal and type II perturbations, respectively.

\section{Conclusion}
\label{sec:conclusion}
In this paper, we have proven that in the long-wavelength limit, we can interpret the Shibata-Sasaki compaction ${\mathscr C}_{\rm SS}$ as the compactness $\tilde{C}$ in the unphysical static spacetime with a metric obtained by removing the cosmological scale factor from the physical spacetime metric, which we call the conformal compactness. Thus, the Shibata-Sasaki compaction can identify circular photon orbits in the unphysical static spacetime that correspond to extremal surfaces and bifurcating trapping horizons in the same spacetime with ${\mathscr C}_{\rm SS}=1/2$, whereas the threshold for PBH formation is given by ${\mathscr C}_{\rm SS}\simeq 0.4$.

The geometrical interpretation proposed here enables one to clearly understand why the Shibata-Sasaki compaction is prohibited to go beyond $1/2$, why
the value of $1/2$ gives an extremal surface on the constant time spacelike hypersurface in the physical spacetime, why it has such a peculiar behaviour 
for the type II configuration and what its value stands for from a geometrical point of view.

In the unphysical static spacetime,
the marginal configuration against the appearance of a circular photon orbit is at the threshold between types I and II of perturbation, while for type II, there are two (or more) extremal surfaces and bifurcating trapping horizons with circular photon orbits. In the physical spacetime, this implies that each extremal surface on the constant time hypersurface has photons that are staying on the surface, as far as the Universe expands in the long-wavelength regime.
In other words, both the maximal and minimal surfaces in the type II 
perturbation behave as expanding photon spheres.

The conformal compactness is conceptually simple, straightforward to calculate, and applicable to spacetimes with any matter contents and/or in modified theories of gravity. If we can neglect the cosmological expansion, which is the case for subhorizon scales, the conformal compactness trivially approaches the compactness, which rigorously identifies future trapping horizons in the physical spacetimes describing gravitational collapse.
To determine the fate of the evolution after the perturbation enters the horizon needs deeper understanding of the general relativistic nonlinear dynamics of the perturbation.

Here, we discuss the status of the Shibata-Sasaki compaction ${\mathscr C}_{\rm SS}$ in comparison to the legitimate compaction in the comoving slice ${\mathscr C}_{\rm com}$. As we have seen, the former is explicitly embedded in the metric function and provides a clear threshold for type II configurations in long-wavelength solutions, while the relationship between the two is not straightforward in general cases. This suggests that ${\mathscr C}_{\rm SS}$ is more fundamental than 
${\mathscr C}_{\rm com}$ from a geometrical perspective. Finally, we should note that the spatial average of the compaction has been reported to be far more reliable than its raw value when the profile dependence of the threshold is considered for PBH formation in 
Ref.~\cite{Escriva:2019phb}. We speculate that this could also be understood from a geometrical viewpoint in the near future.

\acknowledgements

The authors are grateful to C. Germani, M. Kimura, E. A. Lim, H. Maeda,  R.V. Sheth, K. Uehara and M. Yamaguchi for helpful discussion.
T. H. is grateful to CENTRA, Departamento de F\'{i}sica, 
Instituto Superior T\'{e}cnico -- IST at Universidade de Lisboa, and 
Niels Bohr International Academy at Niels Bohr Institute
for their hospitality during the writing of the manuscript.
This work was partially supported by JSPS KAKENHI Grant No. 
JP20H05853 (T. H. , C. Y.), No. JP24K07027 (T. H., C. Y.), No. JP20H05850 (C.Y.), No. JP21K20367 (Y.K.) and No. JP23KK0048 (Y.K.).

\appendix
\section{Stability of the expanding photon spheres as photon surfaces \label{sec:stability}}
We have found that, at a radius where ${\mathscr C}_{\rm SS}=1/2$, there exists a circular null geodesic in the unphysical static spacetime. The static and spherically symmetric timelike hypersurface given by this radius corresponds to a photon sphere.
We also found that the stability of the circular orbit is stable (unstable) if this surface corresponds to the maximal (minimal) surface on the constant time spacelike hypersurface.
This photon sphere is mapped by the conformal transformation to an expanding spherically symmetric timelike hypersurface in the unphysical spacetime, which is a photon surface in more general terminology~\cite{Claudel:2000yi}.
Here we make a remark on this point.

A photon surface is defined as a timelike hypersurface on which there is a tangential photon orbit in every tangential null direction.
It is said to be stable (unstable) if a perturbed photon orbit from the surface is attracted to (repelled from) it~\cite{Koga:2019uqd}.
This behaviour is read off from the sign of a particular component of the Riemann tensor
\begin{align}
    R_{knkn}:=R_{\mu\nu\rho\sigma}k^{\mu}n^{\nu}k^{\rho}n^{\sigma}>0\quad (<0)\quad\Leftrightarrow\quad\mathrm{stable}\quad(\mathrm{unstable}),
\end{align}
where $k^{\mu}$ is the null geodesic tangent and $n^{\mu}$ is the unit normal to the photon surface.
This quantity measures the second-order derivative of the proper length between the perturbed and unperturbed photon orbits.
In a static and spherically symmetric case, this notion of stability is consistent with the standard stability analysis in terms of the second-order derivative of the effective potential for the radial motion of null geodesics.
That is, $R_{knkn}>0$ $(<0)$ at a local minimum (maximum) of the potential.
See Appendix of \cite{Koga:2019uqd} for the proof provided that the areal radius is monotonic and thus can be used as the radial coordinate. However, this proof does not apply for the present case, where the areal radial coordinate encounters coordinate singularity at the photon sphere. So, we first prove below that the stability of the photon sphere corresponds to the second-order derivative of the effective potential even in the present case.

Let us first evaluate the quantity on the photon sphere in the unphysical static spacetime~\eqref{eq:unphysical_static_spacetime}.
The unit normal vector to the surface of $r=r_p$ is given as $\tilde n^\mu=(0,\Psi^{-2},0,0)$.
Because of spherical symmetry, we can choose the tangent null vector as $\tilde k^\mu=(1,0,0,(\Psi^2r\sin\theta)^{-1})$ without loss of generality.
Using Eqs.~\eqref{eq:circular_orbit} and (\ref{eq:V''}), the component of the Riemann tensor for the unphysical static metric is obtained as
\begin{align}
\label{eq:tilde_Rknkn}
    \tilde R_{\tilde k\tilde n\tilde k\tilde n}:=\tilde R_{\mu\nu\rho\sigma}{\tilde k^\mu\tilde n^\nu\tilde k^\rho\tilde n^\sigma}=\frac{3\Psi(r_p) -4\Psi''(r_p) r_{p}^2}{2r^2\Psi^5(r_p)}=-\frac{\tilde{R}''(r_p)}{r_p\Psi^{6}(r_p)}=\frac{1}{2}V''(r_{p}).
\end{align}
Although the unphysical spacetime has a photon sphere at an extremal surface, at which the areal radius is not monotonic, the above discussion proves that the standard stability in terms of the effective potential for the appropriate radial coordinate is equivalent to the stability of the photon sphere in the sense of Ref.~\cite{Koga:2019uqd}.

We are now ready to move onto the stability analysis of the expanding photon sphere or the photon surface in the physical expanding spacetime.
We take the vectors as $n^\mu=a^{-1}\tilde n^\mu$ and $k^\mu=a^{-2}\tilde k^\mu$ so that they are properly normalised with respect to the physical metric.
The component of the Riemann curvature tensor is then obtained as
\begin{align}
\label{eq:Rknkn}
    R_{knkn}
    =a^{-4}\left[\frac{3\Psi(r_p) -4\Psi''(r_p) r^2}{2r^2\Psi^5(r_p)}
    +(\dot a^2-a\ddot a)\right],
\end{align}
where the dot denotes the differentiation with respect to the cosmological time $t$.
By using Eq.~\eqref{eq:tilde_Rknkn} and the Friedmann equation for the flat case, it is rewritten as
\begin{align}
\label{eq:Rknkn_rhop}
    R_{knkn}
    =a^{-4}\left[\tilde R_{\tilde k\tilde n\tilde k\tilde n}
    +(\dot a^2-a\ddot a)\right]
    =a^{-4}\left[\frac{1}{2}V''(r_{p})
    +4\pi G a^2\left(\rho_b+p_b\right)\right],
\end{align}
where $\rho_{b}$ and $p_{b}$ are the energy density and the pressure of the background FLRW solution.
The stability can be changed if the second term in the square brackets dominates the first term.
Interestingly, if we assume the null energy condition $\rho_b+p_b\ge0$, the second term is non-negative and thus has the potential to stabilise the photon surface: the stable photon sphere in the unphysical static spacetime remains stable, whereas the unstable photon sphere in the unphysical static spacetime can become stable in the physical spacetime if the second term dominates the first term.

\bibliographystyle{apsrev4-1}
\bibliography{ref}

\end{document}